\documentclass[aps,prd,11pt]{revtex4-1}
\usepackage{amsmath,amssymb,graphicx,braket,hyperref,bm}
\usepackage{multirow}
\usepackage{float}
\begin{document}

\title{
Adiabatic continuity in a partially reduced twisted Eguchi-Kawai model 
\\with one adjoint Dirac fermion
}

\author{Yudai Hamada}
\email{2644310101e@kindai.ac.jp}
\address{Department of Physics, Kindai University, Osaka 577-8502, Japan}

\author{Tatsuhiro Misumi}
\email{misumi@phys.kindai.ac.jp}
\address{Department of Physics, Kindai University, Osaka 577-8502, Japan}

\begin{abstract}
We numerically investigate whether the center-symmetric confined phase of large-$N$ $SU(N)$ gauge theory with one adjoint Dirac fermion persists under spatial compactification on $\mathbb{R}^3 \times S^1$. To this end, we employ a partially reduced twisted Eguchi-Kawai (TEK) model on a $1^3 \times L_4$ lattice with an adjoint Wilson fermion, and measure both the Polyakov loop around $S^1$ and order parameters for volume independence in the reduced directions. For $N=36$, $L_4=2$, $b=0.30\text{--}0.46$, and $\kappa=0.03\text{--}0.16$, we find that, with periodic boundary conditions, the Polyakov loop remains near zero in the light-fermion regime as the circle size is reduced. For the modified twist, the volume-independence order parameters are also consistent with zero in the explored region, supporting the validity of the partially reduced description. These results provide numerical evidence, within the reduced-model setup and parameter range studied, for an adiabatic-continuity scenario in which the confined phase is smoothly connected between large and small circles. By contrast, with antiperiodic boundary conditions, the Polyakov loop exhibits a clear deconfinement transition. We also discuss how this scenario is compatible with the anomaly constraints of the underlying four-dimensional theory. The symmetric twist is examined as a useful comparison, although its volume-independence properties appear less robust at the present value of $N$.
\end{abstract}

\maketitle
\tableofcontents
\newpage

\section{Introduction}

Volume independence in large-$N$ gauge theory \cite{tHooft:1973alw,Eguchi:1982nm} implies that the physics of $SU(N)$ gauge theory in the $N\to\infty$ limit does not depend on the size of the space-time manifold. In particular, it allows one to reduce the four-dimensional lattice model to a much smaller (even one-site) model without losing information about Wilson loops and other observables, provided the center symmetry remains unbroken.  The simplest one-site reduction of four-dimensional $SU(N)$ lattice gauge theory is the Eguchi-Kawai (EK) model \cite{Eguchi:1982nm}, whose action is 
\begin{equation}
S_{\rm EK} = -bN \sum_{\mu\neq\nu=1}^4 \mathrm{Tr}\Big( U_\mu U_\nu U_\mu^\dagger U_\nu^\dagger \Big)\,,
\end{equation}
with $b=1/(g^2N)$ being the inverse 't Hooft coupling and $U_\mu$ being the link variable.  Volume independence requires the $(\mathbb{Z}_N)^4$ center symmetry under $U_\mu\to z_\mu U_\mu$ ($z_\mu\in\mathbb{Z}_N$) to be preserved. However, the original EK model was found to break this center symmetry at weak coupling \cite{Bhanot:1982sh}: beyond a critical $b$ the Polyakov loops exhibit nonzero expectation values, invalidating the reduction.  

One of the resolutions to the problem \cite{Bhanot:1982sh,Gonzalez-Arroyo:1982hwr, Gonzalez-Arroyo:1982hyq, Narayanan:2003fc, Bringoltz:2009kb, Bringoltz:2011by} is to introduce a ``twist'' phase in the plaquette terms, leading to the twisted Eguchi-Kawai (TEK) model \cite{Gonzalez-Arroyo:1982hwr,Gonzalez-Arroyo:1982hyq}.  In the TEK model the action becomes
\begin{equation}
S_{\rm TEK} = -bN \sum_{\mu\neq\nu=1}^4 \mathrm{Tr}\Big( z_{\mu\nu} \,U_\mu U_\nu U_\mu^\dagger U_\nu^\dagger \Big)\,,
\end{equation}
where $z_{\mu\nu}=\exp(2\pi i k/L) = z_{\nu\mu}^{*}$ are $\mathbb{Z}_N$ phases with $N=L^2$. The twist induces a background flux that frustrates the alignment of Polyakov lines, thereby helping to preserve the $(\mathbb{Z}_N)^4$ center symmetry.  Indeed, for suitably chosen $(k,L)$ (typically $k$ and $L$ coprime, with $k/L$ fixed as $N\to\infty$), the TEK model can maintain the center symmetry and reproduce large-$N$ Yang-Mills results, such as the string tension and meson spectrum \cite{Gonzalez-Arroyo:2012euf,Gonzalez-Arroyo:2015bya,Perez:2020vbn,Bonanno:2023ypf}.  A caveat is that one needs to tune $k$ so that $k/L$ stays above some threshold to avoid the center symmetry breaking in the large-$N$ limit \cite{Gonzalez-Arroyo:2010omx, Guralnik:2002ru, Bietenholz:2006cz, Teper:2006sp}. The possibility of spontaneous breaking of the center symmetry in the TEK model with a simple symmetric twist was first discussed through perturbative analysis in \cite{Guralnik:2002ru}, and subsequently investigated numerically in \cite{Bietenholz:2006cz}. These works considered a partially reduced TEK model, in which two out of the four dimensions are reduced. Later, similar behavior was also confirmed in the fully reduced TEK model, where all directions are reduced \cite{Teper:2006sp}. 

The other approach is to include adjoint fermions in the reduced model \cite{Bringoltz:2009kb, Bringoltz:2011by}.  In the adjoint Eguchi-Kawai (AEK) model one adds $N_f$ adjoint Wilson fermions to the one-site Eguchi-Kawai model.  Since adjoint fermions do not carry a center charge, they force an effective potential to favor a center-symmetric vacuum \cite{Bringoltz:2009kb}.  For example, it was shown that even a single adjoint flavor can help stabilize the $(\mathbb{Z}_N)^4$ symmetry in certain regions of coupling and mass \cite{Bringoltz:2009kb,Bringoltz:2011by}.  However, without a twist, one still faces the problem that at large $b$ and heavy mass the center will break unless the fermion mass is tuned near the critical (massless) point. Furthermore, one cannot calculate physical observables since there is no notion of spatial separation (no emergent space).

Combining both ideas --the twist and adjoint fermions-- yields the adjoint TEK model \cite{Azeyanagi:2010ne, Gonzalez-Arroyo:2012ztz, Gonzalez-Arroyo:2013bta, Gonzalez-Arroyo:2013gpa, GarciaPerez:2013dgk, GarciaPerez:2015rda}, which can further extend the center-symmetric region. 
Indeed, in our previous work using two adjoint flavors \cite{Hamada:2025whg}, we showed that even for the smallest twist $k=1$, heavy adjoint quarks can maintain $(\mathbb{Z}_N)^4$ symmetry without adjusting $k/L$. This allows equivalence to large $N$ gauge theories over a wider parameter range. We also tested the adiabatic continuity of the confinement phase in SU(N) gauge theories on ${\mathbb R}^3 \times S^1$ with two-flavor periodic adjoint fermions by using a partially reduced version of the TEK model with adjoint Wilson fermions. However, the adjoint QCD with $N_f=2$ could potentially exhibit conformal behavior in the massless limit, preventing clear results from being obtained for the confinement phase continuity. Therefore, in this paper, we focus on the case of $N_f=1$. Since the single adjoint flavor has no perturbative infrared fixed point (the two-loop $\beta$ function remains positive at $N_f=1$), the theory is expected to be confined without approaching the conformal phase as the mass is lowered. We ask whether the confined (center-symmetric) phase exhibits adiabatic continuity on $\mathbb{R}^3\times S^1$.

In this paper we study a partially reduced TEK model on a $1^3 \times L_4$ lattice with one adjoint Wilson-Dirac fermion. In this setup, the equivalence to the large-$N$ gauge theory on $\mathbb{R}^3 \times S^1$, or equivalently the volume independence requires that the $(Z_N)^3$ center symmetry in the reduced directions remain unbroken. We therefore examine both the Polyakov loop around $S^1$ and the order parameters relevant for volume independence. Our numerical results at $N=36$ and $L_4=2$ show that, in the light-fermion regime with periodic boundary conditions, the Polyakov loop stays very small throughout the explored range of $b$. For the twist, called ``the modified twist" \cite{Gocksch:1983iw,Gocksch:1983jj}, the volume-independence order parameters are also consistent with zero, which provides numerical evidence, within the reduced-model setup and in the explored parameter region, for {\it adiabatic continuity} of the confined phase \cite{Davies:1999uw, Davies:2000nw, Unsal:2007vu, Unsal:2007jx, Kovtun:2007py, Shifman:2008ja, Unsal:2010qh, Poppitz:2009uq, Anber:2011de, Poppitz:2012sw, Misumi:2014raa, Yamazaki:2017ulc,Shimizu:2017asf, Tanizaki:2017qhf, Tanizaki:2017mtm, Hongo:2018rpy,Misumi:2019dwq,Fujimori:2019skd, Misumi:2019upg, Fujimori:2020zka, Unsal:2020yeh, Poppitz:2021cxe, Tanizaki:2022ngt, Tanizaki:2022plm, Hayashi:2023wwi, Hayashi:2024qkm, Hayashi:2024gxv, Hayashi:2024yjc, Hayashi:2024yjc, Hayashi:2024psa, Guvendik:2024umd,Anber:2024mco, Hayashi:2025odr, Hayashi:2025mgk}: the confined, center-symmetric phase at large $S^1$ radius persists to small radius with no phase transition. For comparison, we also present results for the symmetric twist \cite{Gonzalez-Arroyo:2010omx}, whose volume-independence properties appear less robust at this value of $N$. By contrast, with antiperiodic (thermal) boundary conditions, we observe the expected deconfinement transition.

The rest of this paper is organized as follows.  In Sec.~\ref{sec:TEK} we define the one-flavor adjoint TEK model. In Sec.~\ref{sec:FT} we introduce the partially reduced $\mathbb{R}^3 \times S^1$ setup and define the order parameters relevant for volume independence and confinement, where we numerically examine the volume independence for the symmetric and modified twists. In Sec.~\ref{sec:AC} we present our numerical results for adiabatic continuity. In Sec.~\ref{sec:anomaly} we discuss how the adiabatic-continuity scenario is constrained by the 't Hooft anomaly of the underlying theory. Sec.~\ref{sec:SD} contains a summary and discussion.

\section{Adjoint Twisted Eguchi-Kawai model} 
\label{sec:TEK}

The twisted Eguchi-Kawai (TEK) one-site model for $SU(N)$ gauge theory in four dimensions is defined by the action
\begin{equation}
S_{\rm TEK} = -bN \sum_{\mu\neq\nu=1}^4 \mathrm{Tr}\Big( z_{\mu\nu}\,U_\mu U_\nu U_\mu^\dagger U_\nu^\dagger \Big)\,,
\end{equation}
where $U_\mu$ ($\mu=1,\dots,4$) are $SU(N)$ link variables on the single site, and $z_{\mu\nu}\in\mathbb{Z}_N$ are constant twist phases.  A convenient choice called the symmetric twist is 
\begin{equation}
z_{12}=z_{23}=z_{34}=z_{41}= e^{2\pi i k/L}, \quad
z_{21}=z_{32}=z_{43}=z_{14}= e^{-2\pi i k/L}, \quad N=L^2, 
\label{eq:twist}
\end{equation}
with other $z_{\mu\nu}=1$.  
As discussed in the introduction, $k$ and $L$ are required to be coprime, with $k/L$ being fixed as $N\to\infty$.
One of the proposals is to set $k/L$ to a ratio of every-other Fibonacci numbers as $\displaystyle \sim \lim_{j\to \infty} F_{j} /F_{j+2} \to 0.382$ \cite{Chamizo:2016msz,Hayashi:2025doq}.

We now add a single flavor ($N_f=1$) of adjoint Wilson-Dirac fermion to the one-site lattice.  We impose periodic boundary conditions for the adjoint fermion in all directions, which is appropriate for preserving center symmetry.  The fermion action on the single site is 
\begin{equation}
S_F = \bar\psi\,D_W[U]\,\psi,
\end{equation}
where $D_W$ is the Wilson-Dirac operator in the adjoint representation, given by
\begin{equation}
D_W[U] = 1 - \kappa \sum_{\mu=1}^4\big[(1-\gamma_\mu)U_\mu^{\rm adj} +(1+\gamma_\mu)U_\mu^{{\rm adj},\dagger}\big]\!,
\label{eq:Dw}
\end{equation}
with $\kappa$ the hopping parameter.  Here $U_\mu^{\rm adj}$ are the $(N^2 -1 )\times (N^2 -1)$ matrices in the adjoint representation, $(U_\mu^{\rm adj})_{ab} = 2\,\mathrm{Tr}(T_a U_\mu T_b U_\mu^\dagger)$ with $T_a$ the $SU(N)$ generators.  The fermion mass $m_0$ is related to $\kappa$ by $m_0=1/(2\kappa)-1/(2\kappa_c)$, where $\kappa_c$ is the critical hopping parameter corresponding to the massless limit. In this work, we focus on the light-mass regime, where the adjoint quark can significantly affect the dynamics. By tuning $\kappa$ close to $\kappa_c$, we consider a situation in which the adjoint quark becomes light and contributes significantly as a dynamical degree of freedom.

The combined action of the adjoint TEK model is then 
\begin{align}
S_{\rm ATEK} = S_{\rm TEK} \;+\; \bar\psi \,D_W[U]\,\psi .
\end{align}

Adding the adjoint fermion does not change the gauge twist structure, but it does influence the center-symmetry.  Since adjoint matter carries no net center charge, a nonzero expectation value of a Polyakov loop in one direction increases the fermionic free energy.  In effect, the adjoint fermion exerts a ``restoring force'' on any broken center configuration \cite{Bringoltz:2009kb,Bringoltz:2011by}.  Indeed, our previous study has shown that adjoint fermions can stabilize $(\mathbb{Z}_N)^4$ symmetry \cite{Hamada:2025whg}.  

It is also useful to comment on the continuum dynamics: $SU(N)$ gauge theory with one adjoint fermion is asymptotically free ($b_0>0$) and has no perturbative infrared fixed point ($b_1>0$ at two-loop).  Thus $N_f=1$ adjoint QCD is expected to be confining at all scales, unlike the $N_f=2$ case which is near the conformal window \cite{DelDebbio:2010zz,Athenodorou:2024rba}.

\section{Partially reduced model on ${\mathbb R}^3 \times S^1$}
\label{sec:FT}

\subsection{Definition and order parameters}
\label{sec:PRM}

 Consider $SU(N)$ gauge theory on $\mathbb{R}^3 \times S^{1}$, where $S^1$ is a circle of circumference ${\mathcal L}_{4}$. In the $N\to\infty$ limit, a reduced model corresponding to this theory can be formulated as a modified version of the TEK model. As in our previous work \cite{Hamada:2025whg}, we employ a partially reduced TEK model with a lattice of size $1^3 \times L_4$, where three directions $\mu=1,2,3$ are one-site and one direction $\mu=4$ has $L_4$ sites. ${\mathbb Z}_N$ twisted boundary conditions are imposed in the three reduced directions, while standard periodic boundary conditions are imposed in the extended direction. 
The action \cite{Gocksch:1983iw, Gocksch:1983jj, Azeyanagi:2010ne} is 
\begin{align}
S_{{\mathbb R}^3\times S^1}\, &=\, 
-bN\sum_{\tau=0}^{L_4 -1} \mathrm{Tr}\left(  \sum_{i=1}^{3}U_4(\tau) U_i(\tau+1) U_4^\dagger(\tau) U_i^\dagger(\tau)  + \sum_{i\neq j}^3  z_{ij} U_i(\tau) U_j(\tau) U_i^\dagger(\tau) U_j^\dagger(\tau) \right)+S_{F}\,,
\label{eq:R3S1}
\end{align}
where $\tau$ stands for lattice sites in the 4th direction with the size $L_4$ and $U_{4}(\tau)$ is the link variable along this direction while $U_{i}(\tau)$ ($i=1,2,3$) are the link variables along the three reduced directions with one site.
In this paper, we consider $L_4 \ll L=\sqrt{N}$.

We note that in the three directions $i=1,2,3$, an effective $L^3 =N^{3/2}$ lattice effectively ``emerges" in the reduced directions as a consequence of the ${\mathbb Z}_N$ twist while the lattice size in the 4th direction is $L_4$, leading to physical circumference ${\mathcal L}_4 = L_4 a$, with $a$ an effective lattice spacing.
This indicates that the model can describe the lattice gauge theory on $L^3 \times L_4$ with $L_4 \ll L$ \cite{Azeyanagi:2010ne}.
This model corresponds to a finite-temperature system if antiperiodic boundary conditions are imposed for the adjoint fermion in the 4th direction, or to the theory with a spatial compactification if periodic boundary condition is imposed in the direction. We mainly focus on the latter case with periodic adjoint fermions since it is relevant to \emph{adiabatic continuity}. 

In this partially reduced model, the condition for volume independence on $\mathbb{R}^3 \times S^1$, or the condition for the equivalence of the continuum theory and the partially reduced model in the large-$N$ limit, is that $(\mathbb{Z}_N)^3$ center symmetry in the three reduced directions remains unbroken. With this condition satisfied, the model on the $1^3\times L_4$ lattice is equivalent to the lattice gauge theory on $L^3 \times L_4$. 
We define the Wilson-loop expectation value as
\begin{align}
W = \frac{1}{3N L_4} \sum_{\tau=0}^{L_4-1}\sum_{i=1}^3 \left|\langle\mathrm{Tr}U_i(\tau) \rangle\right|\,,
\label{eq:acW}
\end{align}
which is an order parameter for the $(\mathbb{Z}_N)^3$ symmetry for the volume independence. $U_i(\tau)$ are the link variables in the reduced directions, which depend on the coordinate of the extended direction $\tau$.
It is worth noting that the condition $W=0$ is a necessary condition, but not a sufficient condition for the volume independence, and the necessary and sufficient condition for the volume independence is that all the possible Wilson lines defined below are zero for $-\sqrt{N} < n_1, n_2, n_3 < \sqrt{N}$:
\begin{align}
W(n_1, n_2, n_3) \equiv  \frac{1}{N L_4}  \sum_{\tau=0}^{L_4-1} \left|\langle\mathrm{Tr}[(U_1(\tau))^{n_{1}}(U_2(\tau))^{n_{2}}(U_3(\tau))^{n_{3}}] \rangle\right|\,.
\label{eq:wn123}
\end{align}

In our previous work \cite{Hamada:2025whg}, we discussed volume independence in the partially reduced $N_f = 2$ adjoint TEK model on $\mathbb{R}^3 \times S^1$, and found that some of the Wilson lines $W(n_1, n_2, n_3)$ are not consistent with zero at least for small $N$ in the case of the symmetric twist in Eq.(\ref{eq:twist}) while they are consistent with zero in the case of the modified twist, which is defined as 
\begin{align} &z_{12} \,=\, z_{23} \,=\, \exp \left(\frac{2\pi i l}{L} \right),\qquad z_{13} \,=\, \exp \left(\frac{\pi i l}{L} \right), \label{eq:mtwist} 
\end{align} 
with $l$ an integer smaller than $L$, $z_{ji} = z_{ij}^{*}$ ($i=1,2,3$) and $N=L^2$.
Motivated by this result, we consider both types of the twists in the partially reduced model with $N_f =1$ in the present work. 
In Sec.~\ref{sec:VI}, we present calculations that test volume independence in the partially reduced TEK models with the symmetric twist and the modified twist.

On the other hand, the order parameter of the (de)confinement, or the indicator of adiabatic continuity, is the Polyakov loop expectation value around the $S^1$ (the 4th direction)
\begin{align} P = \frac{1}{N}\left|\langle \mathrm{Tr}\prod_{\tau=0}^{L_4 -1} U_4(\tau)\rangle\right|\,. \label{eq:acP} 
\end{align}
This observable diagnoses center-symmetry realization along the compact $S^1$ and is therefore useful for probing a possible deconfinement transition. In Sec.~\ref{sec:AC}, we numerically simulate the $N_f = 1$ partially reduced adjoint TEK model on $1^3 \times L_4$ with $N=36$, $L_4=2$, $b=0.30$-$0.46$, and $\kappa =0.03, 0.16$.
In the simulation with antiperiodic and periodic boundary conditions in the $S^1$ (4th) direction for the adjoint fermion, we will measure $P$ as functions of $b$ for given $\kappa$ and $L_4$ to study whether or not a deconfinement transition occurs, where a larger (smaller) $b$ corresponds to a smaller (larger) circumference. We there use the rational hybrid Monte Carlo method to simulate the model with a single adjoint fermion ($N_f = 1$).

\subsection{Volume independence in partially reduced adjoint TEK models} \label{sec:VI}

Based on the discussion in our previous paper \cite{Gocksch:1983iw, Gocksch:1983jj, Hamada:2025whg}, in order to investigate the issue on volume independence in the partially reduced TEK models with the symmetric and modified twists, it is necessary to compute, in addition to the volume independence order parameter $W$ in Eq.~(\ref{eq:acW}), the following quantity $W_{123}$ \cite{Okawa:2025}:
\begin{align} 
W_{123} = \frac{1}{N L_4} \sum_{\tau=0}^{L_4-1} \left|\langle\mathrm{Tr}(U_1(\tau) U_2^{\dagger}(\tau) U_3(\tau) )\rangle\right|\,, 
\label{eq:w123}
\end{align}
which is also the order parameter of the $(\mathbb{Z}_N)^3$ symmetry. 
In the symmetric twist, the simplest single-link order parameter $W$ is insufficient because a special composite loop $W_{123}$ can remain nonzero even when each elementary link is traceless.
Although the detailed reason why this specific combination is special among all possible $(n_1, n_2, n_3)$ was extensively discussed in Ref.~\cite{Hamada:2025whg}, we briefly review this in Appendix \ref{sec:sol} by examining the classical solutions of the partially reduced model.

In this subsection, we show the numerical Monte Carlo calculations of both $W$ and $W_{123}$ for $N_f=1$, $N=36$, $\kappa=0.03,0.16$, $b=0.34$-$0.46$, $L_4 = 2$ with $k=1$ for the symmetric twist in Eq.~(\ref{eq:twist}) and $l=3$ for the modified twist in Eq.~(\ref{eq:mtwist}). The setup of the simulation is shown in Appendix \ref{sec:SS}.

\begin{figure}[t]
\centering
\includegraphics[width=8cm]{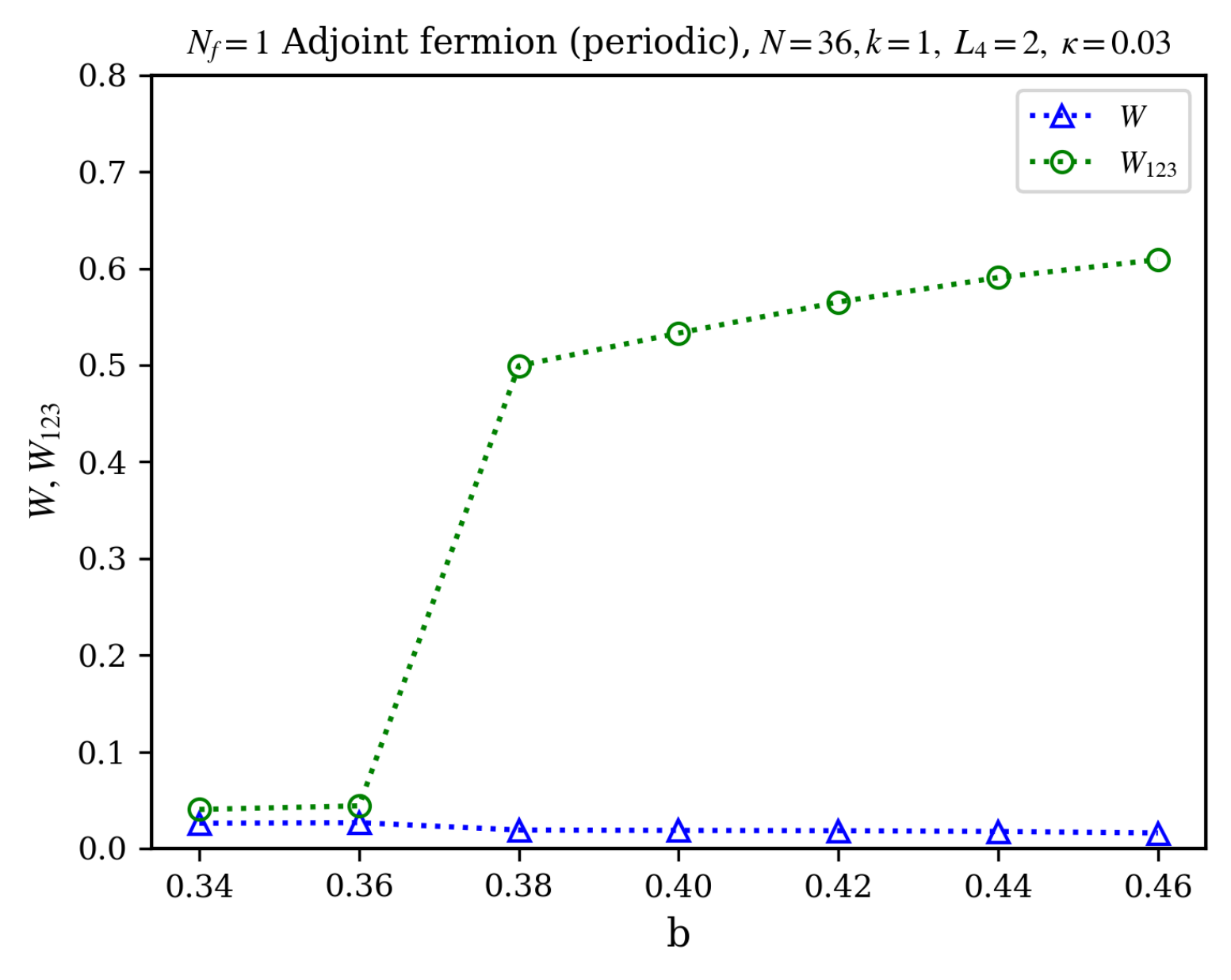}
\includegraphics[width=8cm]{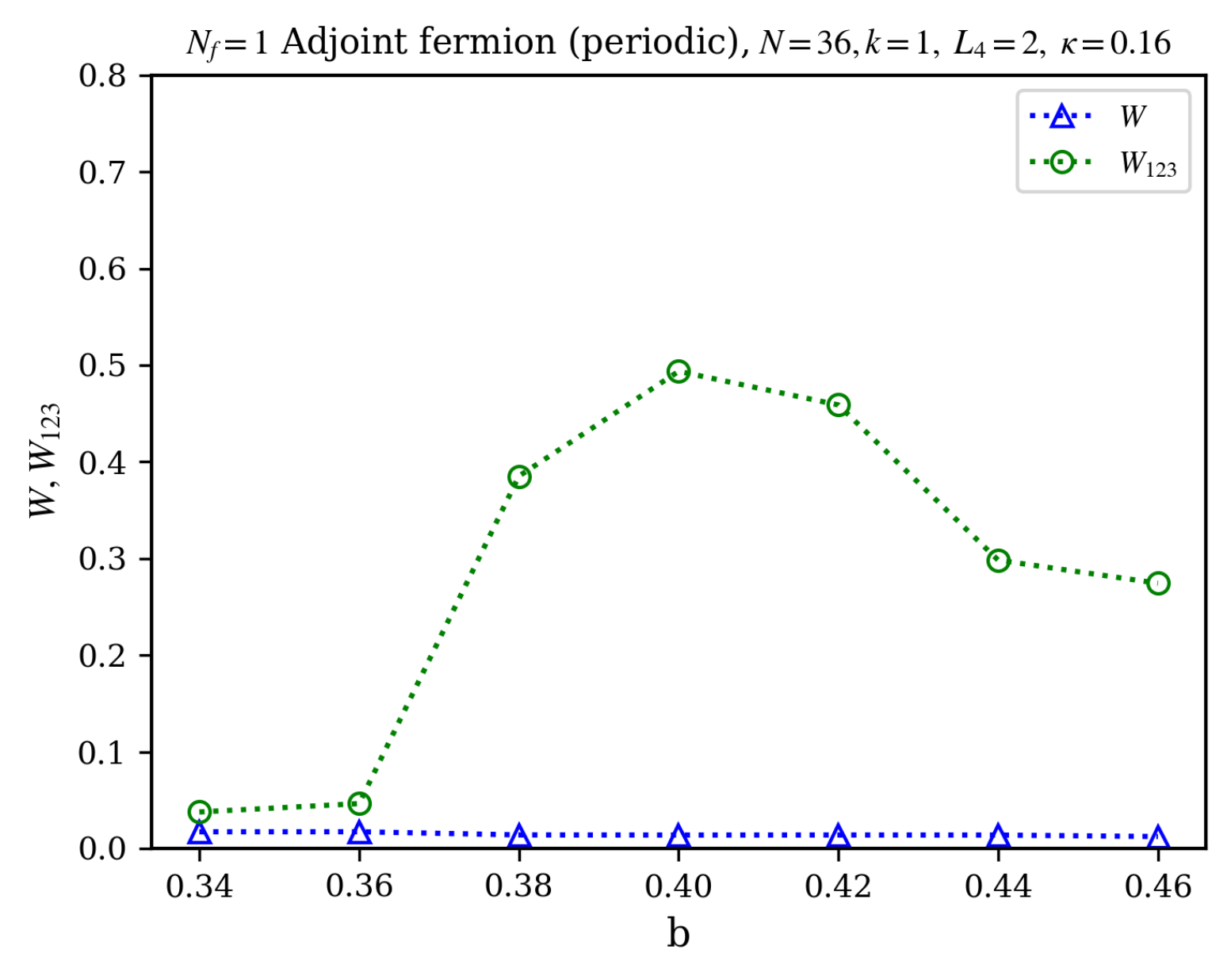}
\caption{The volume-independence order parameters $W, W_{123}$ in the partially reduced TEK model with {\bf the symmetric twist}, $N=36$, $N_f=1$, $k=1$ and $L_4 = 2$ as a function of $b$. (Left) Heavy adjoint fermions $\kappa=0.03$: $W$ is consistent with zero, but $W_{123}$ has nonzero values. (Right) Light adjoint fermion $\kappa=0.16$: $W$ is consistent with zero while $W_{123}$ is nonzero, but gets smaller as $b$ increases. Error bars are smaller than the symbol size.}
\label{fig:VI}
\end{figure}

\begin{figure}[t]
\centering
\includegraphics[width=8cm]{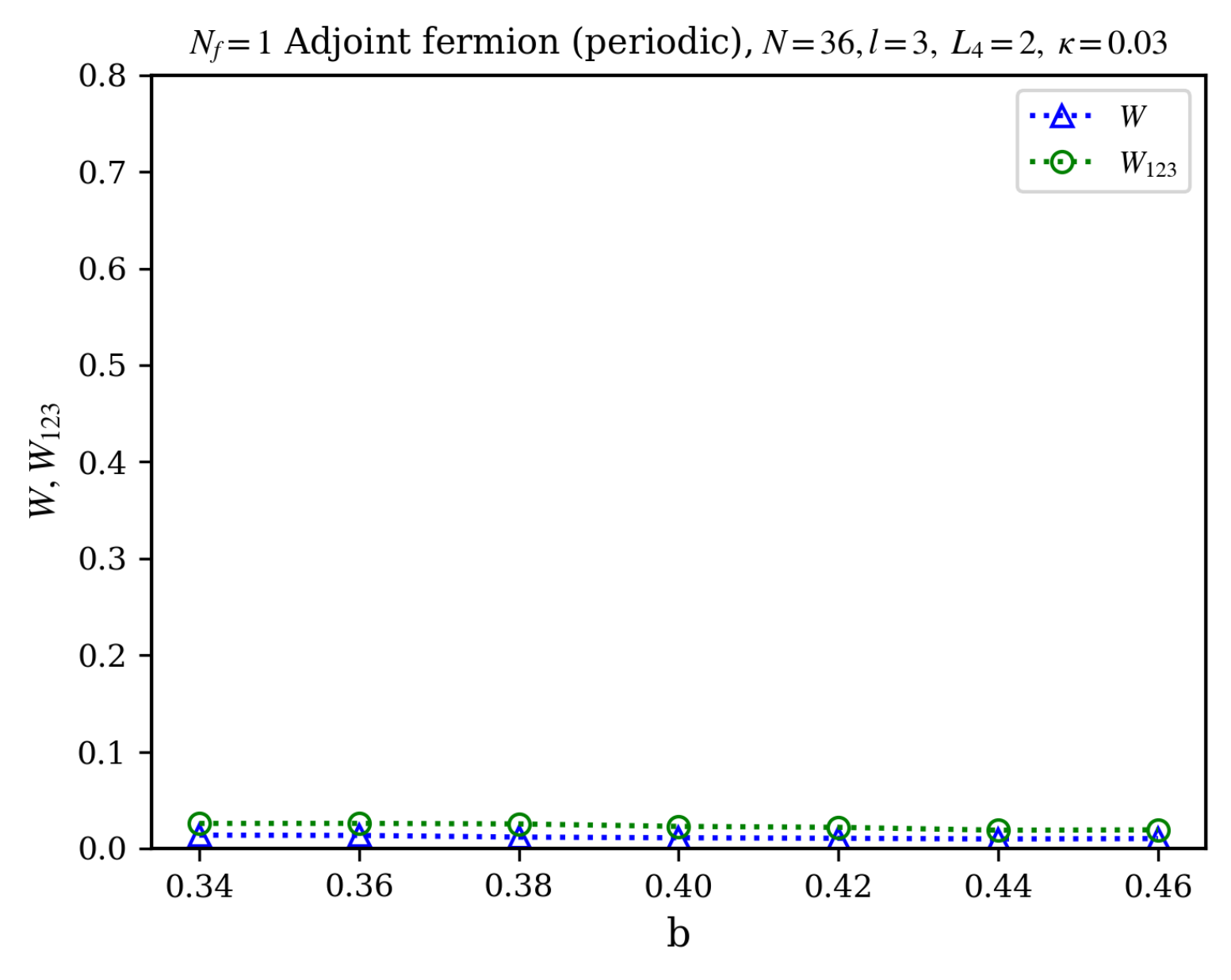}
\includegraphics[width=8cm]{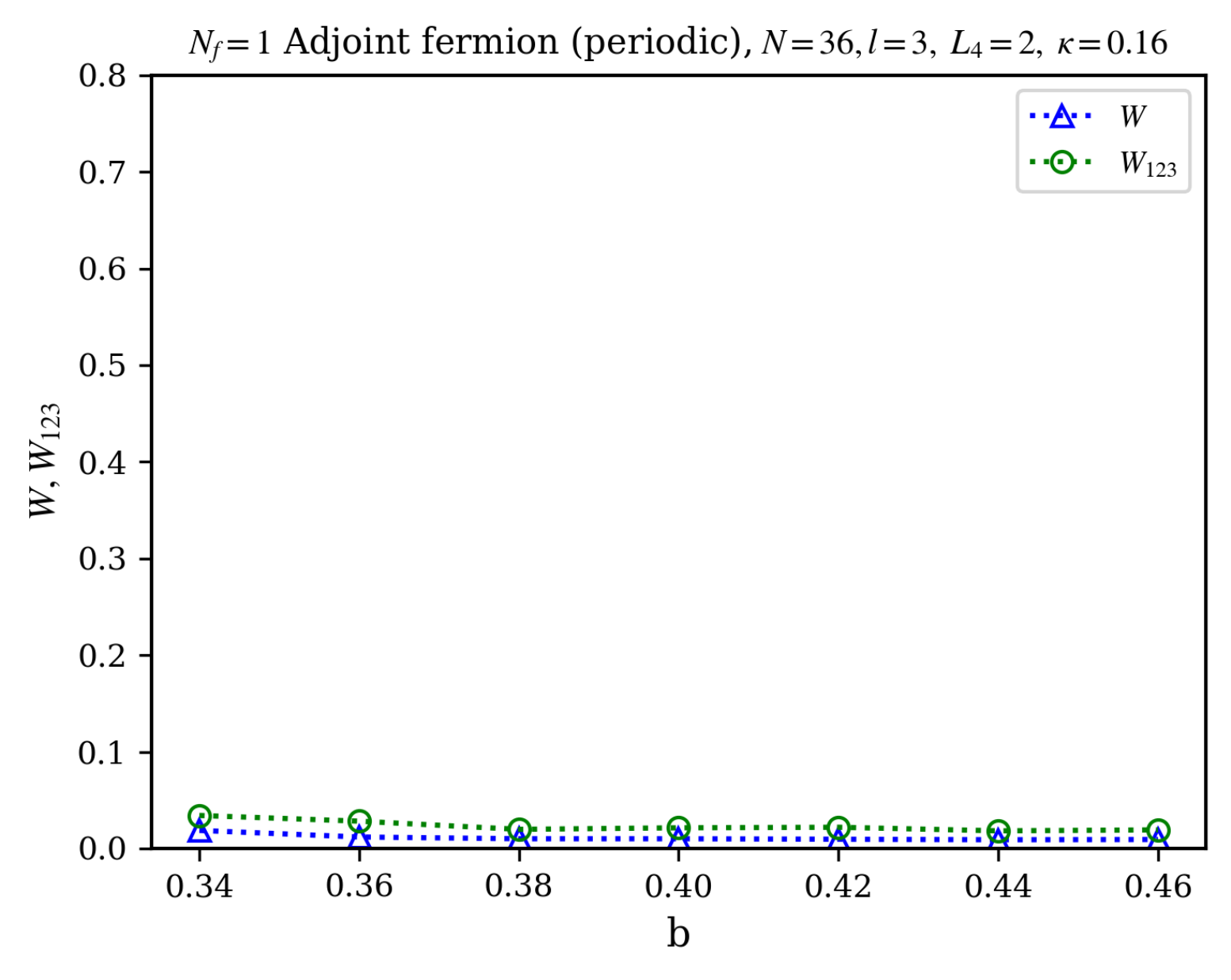}
\caption{The volume-independence order parameters $W, W_{123}$ in the partially reduced TEK model with {\bf the modified twist}, $N=36$, $N_f=1$, $l=3$ and $L_4 = 2$ as a function of $b$. (Left) Heavy adjoint fermions $\kappa=0.03$:  $W, W_{123}$ remains near zero over the entire range. (Right) Light adjoint fermions $\kappa=0.16$: $W, W_{123}$ remains near zero over the entire range. Error bars are smaller than the symbol size.}
\label{fig:VI2}
\end{figure}

Figure \ref{fig:VI} shows the results of $W,\,W_{123}$ for $\kappa = 0.03$ (heavy mass) and $\kappa = 0.16$ (light mass) in the case of the {\bf symmetric twist} defined in Eq.~(\ref{eq:twist}), where the critical $\kappa$ corresponding to a massless limit is $\kappa_c \approx 0.178$.
First of all, $W$ is consistent with zero for both $\kappa = 0.03$ and $\kappa = 0.16$ in the considered region of $b$.
For $\kappa = 0.03$ (left panel), $W_{123}$ takes clearly nonzero values, suggesting that the $(\mathbb{Z}_N)^3$ symmetry may be broken, at least for heavy adjoint fermions. 
On the other hand, for $\kappa = 0.16$ corresponding to light adjoint fermions (right panel), the $(\mathbb{Z}_N)^3$ center symmetry appears to be broken in the intermediate region. However, as $b$ increases, $W_{123}$ is getting small. This behavior suggests the possibility of a partial breaking/restoration of the $(\mathbb{Z}_N)^3$ center symmetry.
Although the results in Fig.~\ref{fig:VI} are derived for small $N$ and we need further investigation to study the behavior of $W, W_{123}$ in the large-$N$ region, at this stage, they lead us to be skeptical about the volume independence in the partially reduced Eguchi-Kawai model with the symmetric twist.

By contrast, the model with the modified twist presents a markedly different picture.
Figure \ref{fig:VI2} shows the results of $W,\,W_{123}$ for $\kappa = 0.03$ (left panel) and $\kappa = 0.16$ (right panel) in the case of the {\bf modified twist} defined in Eq.~(\ref{eq:mtwist}). 
For both $\kappa = 0.03$ and $0.16$, both of $W$ and $W_{123}$ are consistent with zero, suggesting that the $(\mathbb{Z}_N)^3$ symmetry is unbroken in the present region. 
This result indicates that the volume independence in the partially reduced model with the modified twist survives in the broad parameter region, which allows us to investigate the adiabatic continuity of confining phase in the system.


\section{Numerical results for adiabatic continuity}
\label{sec:AC}

\subsection{Deconfinement transition with antiperiodic adjoint fermions}

\begin{figure}[t]
\centering
\includegraphics[width=8cm]{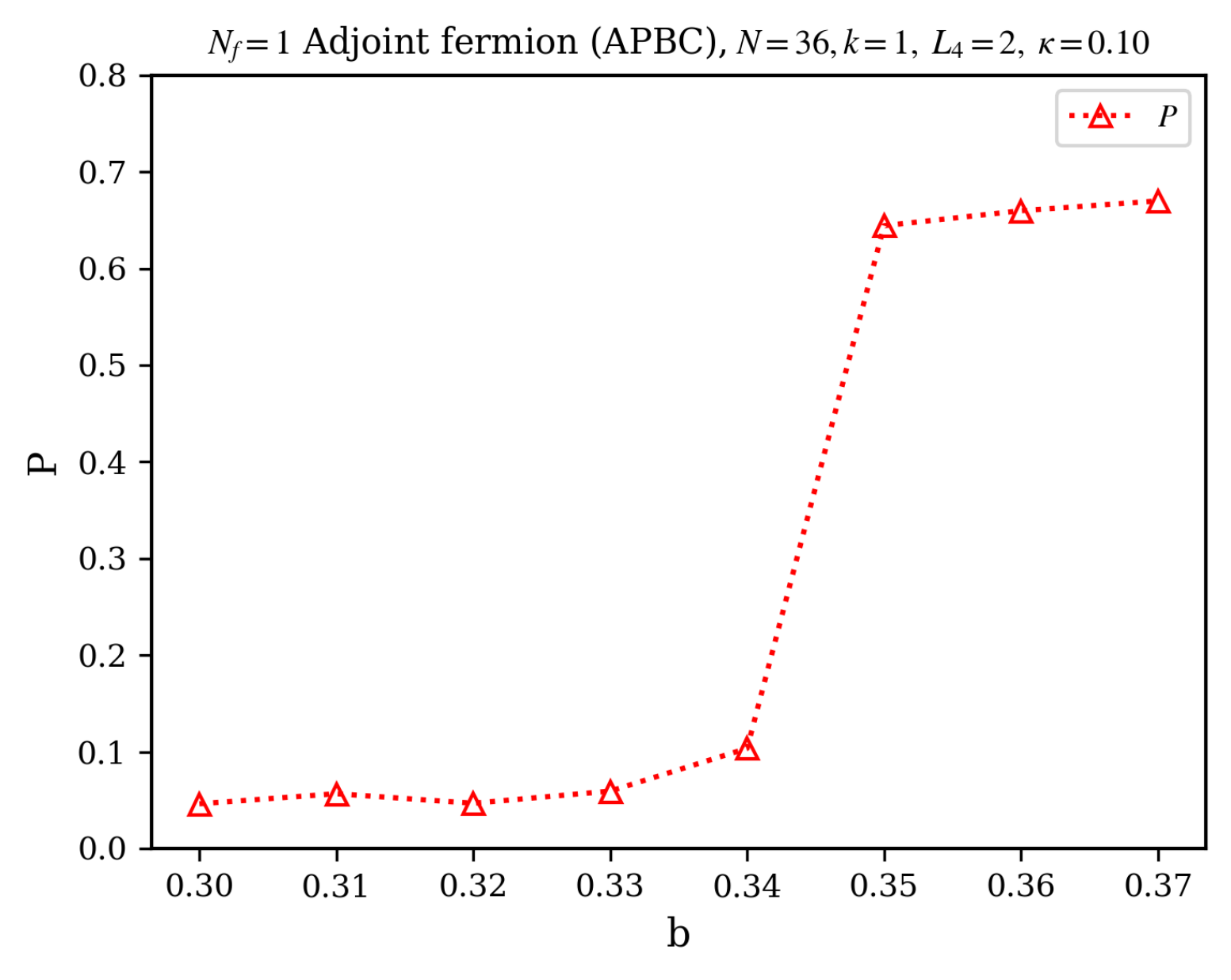}
\includegraphics[width=8cm]{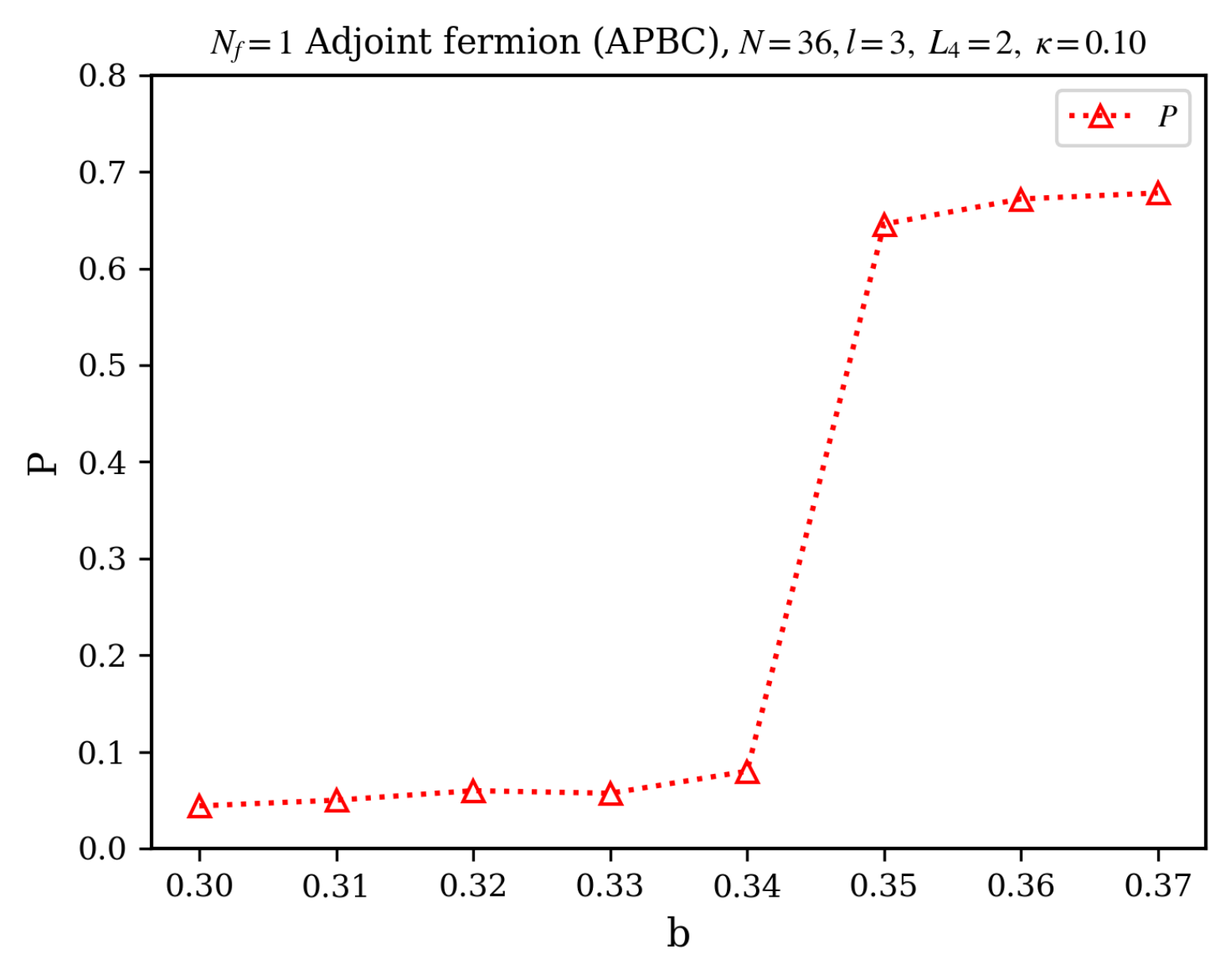}
\caption{Deconfinement order parameter $P$ (red triangles) 
in the partially reduced TEK model with $N_f=1$ antiperiodic adjoint fermions, $N=36$, $\kappa=0.10$ and $L_4 = 2$ as a function of $b$. 
(Left) {\bf symmetric twist} with $k=1$.
(Right) {\bf modified twist} with $l=3$.
Error bars are smaller than the symbol size.}
\label{fig:APBC1}
\end{figure}

We first study the model with a single adjoint flavor with antiperiodic boundary condition in the $S^1$ (4th direction), which corresponds to the finite-temperature system.
In Figure \ref{fig:APBC1}, we show the deconfinement order parameter $P$ in the partially reduced TEK model with a one-flavor antiperiodic adjoint fermion, $N=36$, $L_4 = 2$, $\kappa=0.10$ as a function of $b$. 
In the left panel we show the result for the symmetric twist $k=1$, while in the right panel we show the result for the modified twist $l=3$.
For both twists, the $S^1$ Polyakov loop $P$ shows a sharp rise around $b=0.34$, indicating deconfinement phase transition, consistent with the known results in \cite{Azeyanagi:2010ne, Hamada:2025whg}.
The setup of the simulation is shown in Appendix \ref{sec:SS}.

This thermal case serves as a useful benchmark for the partially reduced model. Namely, it shows that the model reproduces the expected deconfinement behavior when the adjoint fermion obeys antiperiodic boundary conditions. This provides a clear contrast with the periodic case studied in the following subsections, where the absence of such a transition in the light-fermion regime is the key signature of the adiabatic-continuity scenario.

\subsection{Adiabatic continuity with the symmetric twist} \label{sec:ac}

In this and next subsections, we show the numerical results of the deconfinement order parameter $P$ in the partially reduced TEK model with one-flavor adjoint fermion with periodic boundary conditions in the $S^1$ direction (4th direction). 

Although in Sec.~\ref{sec:VI} we have seen the negative results on the volume independence in the model with the symmetric twist Eq.~(\ref{eq:twist}), there is still a possibility that $({\mathbb Z}_N)^3$ symmetry is recovered and the volume independence exists in the large-$N$ limit. 
Thus, we consider it meaningful to discuss the adiabatic continuity in the model with the symmetric twist although the following results are suggestive only.

We show the Polyakov loop $P$ as a function of $b$ in the model with a one-flavor periodic adjoint fermion, $N=36$, $k=1$(symmetric twist), $L_4 = 2$ and $\kappa = 0.03, 0.16$. The setup of the simulation is given in Appendix \ref{sec:SS}.

In Fig.~\ref{fig:AC1}(left panel), with $\kappa = 0.03$ (heavy mass), the quantity $P$ clearly departs from zero at $b \approx 0.36$, signaling a phase transition at small $S^1$.
This demonstrates that if adjoint fermions are too heavy, close to pure gauge theory, the model shows a deconfinement transition at certain critical coupling $b_c$, effectively corresponding to the critical circumference of compactification.

In contrast, Fig.~\ref{fig:AC1}(right panel) with $\kappa=0.16 < \kappa_c \simeq 0.178$ (light mass) shows that $P$ stays very small for all $b$ in the range with no noticeable jump. 
This is consistent with the absence of any transition, i.e. it suggests that the theory remains in the confined phase even when the spatial compactification circumference is quite small. We note that the volume-independence properties of the symmetric twist are less clear at the present value of $N$; thus, this result should be regarded as suggestive rather than conclusive.

\begin{figure}[t]
\centering
\includegraphics[width=8cm]{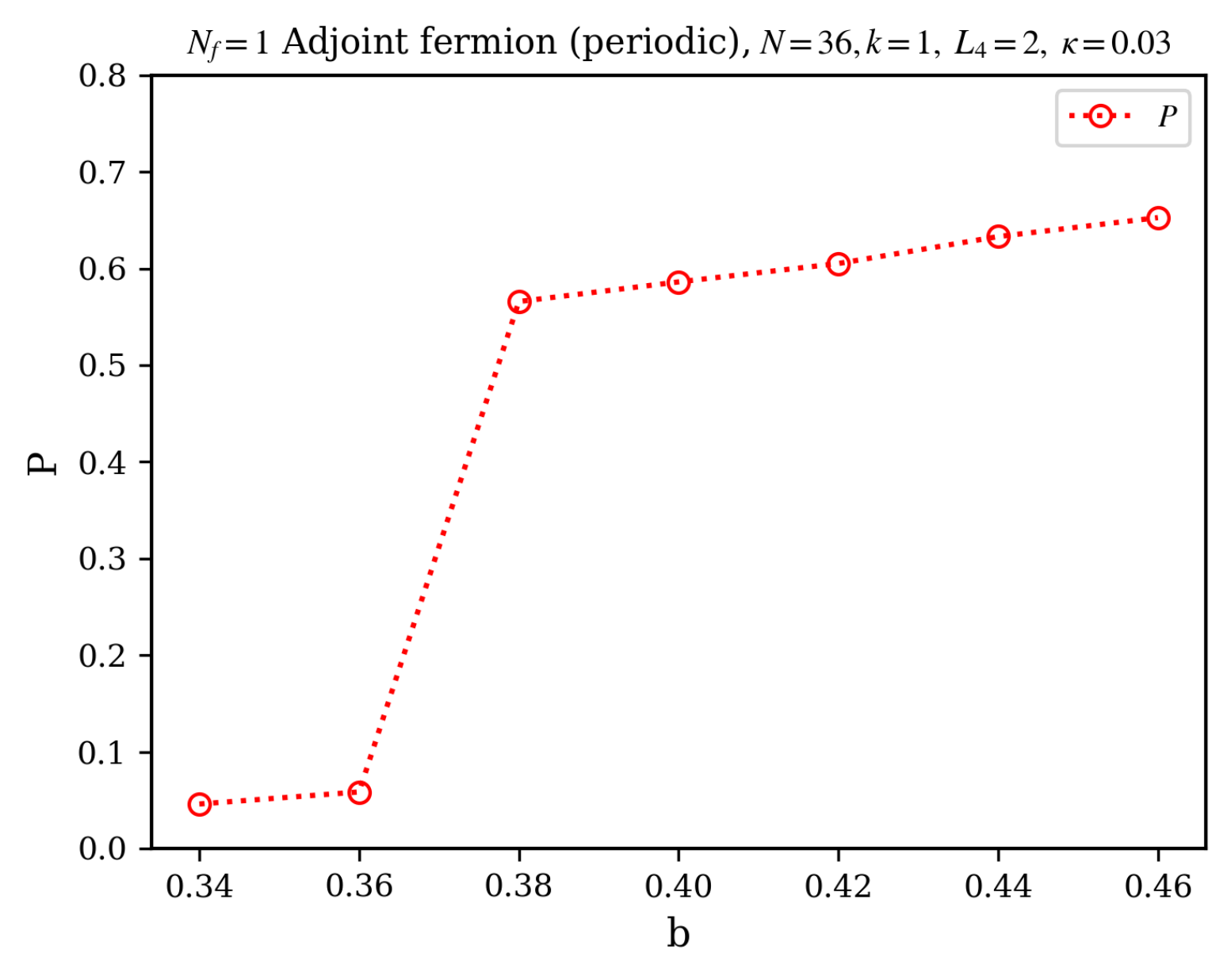}
\includegraphics[width=8cm]{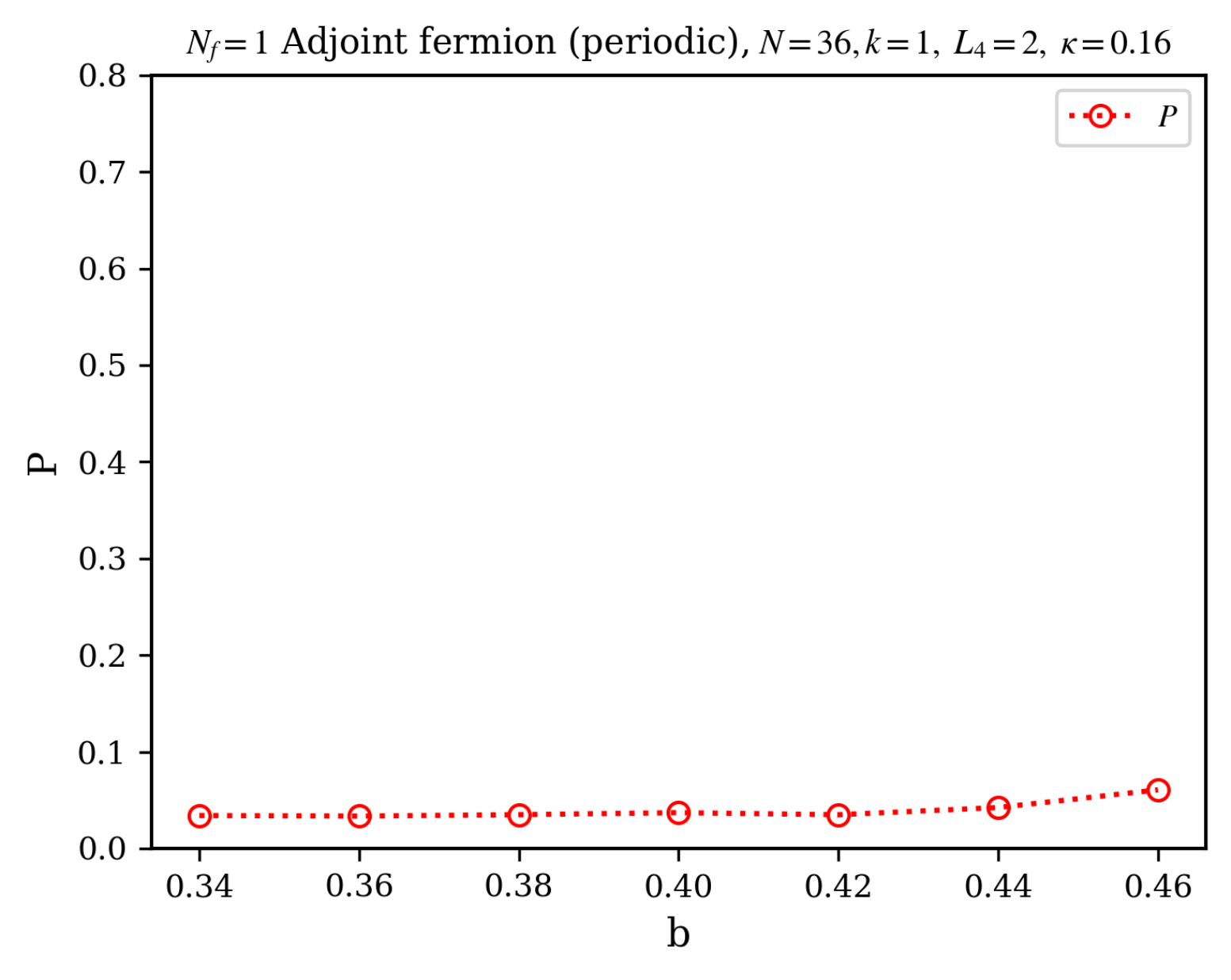}
\caption{Deconfinement order parameter $P$ (red circles) in the partially reduced TEK model with {\bf the symmetric twist}, $N=36$, $N_f=1$ (periodic b.c.), $k=1$ and $L_4 = 2$ as a function of $b$. (Left) Heavy adjoint fermions $\kappa=0.03$:  $P$ shows a sharp rise around $b=0.36$, indicating deconfining phase transition. (Right) Lighter adjoint fermions $\kappa=0.16$: $P$ remains near zero over the entire range, with no sign of a phase transition. Error bars are smaller than the symbol size.}
\label{fig:AC1}
\end{figure}

\subsection{Adiabatic continuity with the modified twist} \label{sec:MT}

In Fig.~\ref{fig:VI2} of Sec~\ref{sec:VI} we show the numerical evidence of the volume independence in the partially reduced model with this modified twist in Eq.~(\ref{eq:mtwist}).
It is  of great importance to investigate the adiabatic continuity of confining phase in the model, where the volume independence is well supported numerically.

Thus, we adopt the modified twist in Eq.~(\ref{eq:mtwist}) in the partially reduced TEK model with a single adjoint fermion with periodic boundary condition in $S^1$ direction.
Here, we consider $N = 36$, and set $l=3$ in the modified twist. It is worth noting that, for large $N$, $l$ must be increased to maintain $(\mathbb{Z}_N)^3$ center symmetry. 
The setup of the simulation is given in Appendix \ref{sec:SS}.

In Fig.~\ref{fig:MT}, we show the numerical results of the deconfinement order parameter $P$ 
with $\kappa=0.03$ (heavy mass) and $\kappa=0.16$ (light mass) in the partially reduced TEK model with the modified twist in Eq.~(\ref{eq:mtwist}), $N=36$, $N_f=1$ (periodic b.c.) and $L_4 = 2$ as a function of $b$.

In the left panel of Fig.~\ref{fig:MT} for $\kappa=0.03$, $P$ clearly departs from zero, signaling the phase transition for the small $S^1$.
In the right panel of Fig.~\ref{fig:MT} for $\kappa=0.16$, there is no jump in $P$ staying very small over the entire range of $b$. This indicates the absence of a transition, suggesting that the theory remains in the confined phase when the compactification circumference is varied from large to small. We note that the small but nonzero values of $P$ for $\kappa = 0.16$ in the right panel of Fig.~\ref{fig:MT} may originate from the small-$N$ effects and will disappear in a large $N$ \cite{Hamada:2025whg}. This provides strong support for adiabatic continuity in the model, together with evidence for volume independence in the explored parameter region.

\begin{figure}[t]
\centering
\includegraphics[width=8cm]{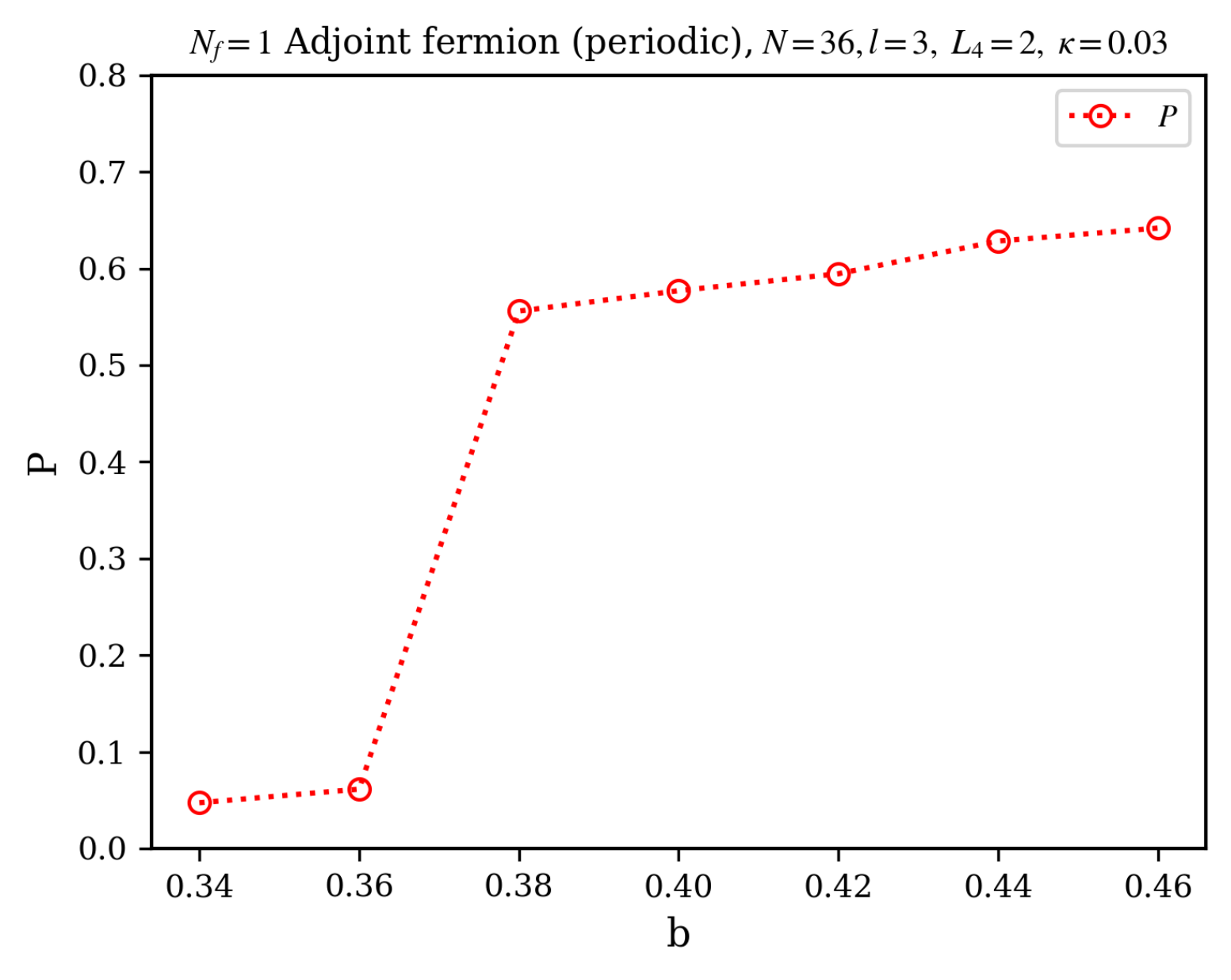}
\includegraphics[width=8cm]{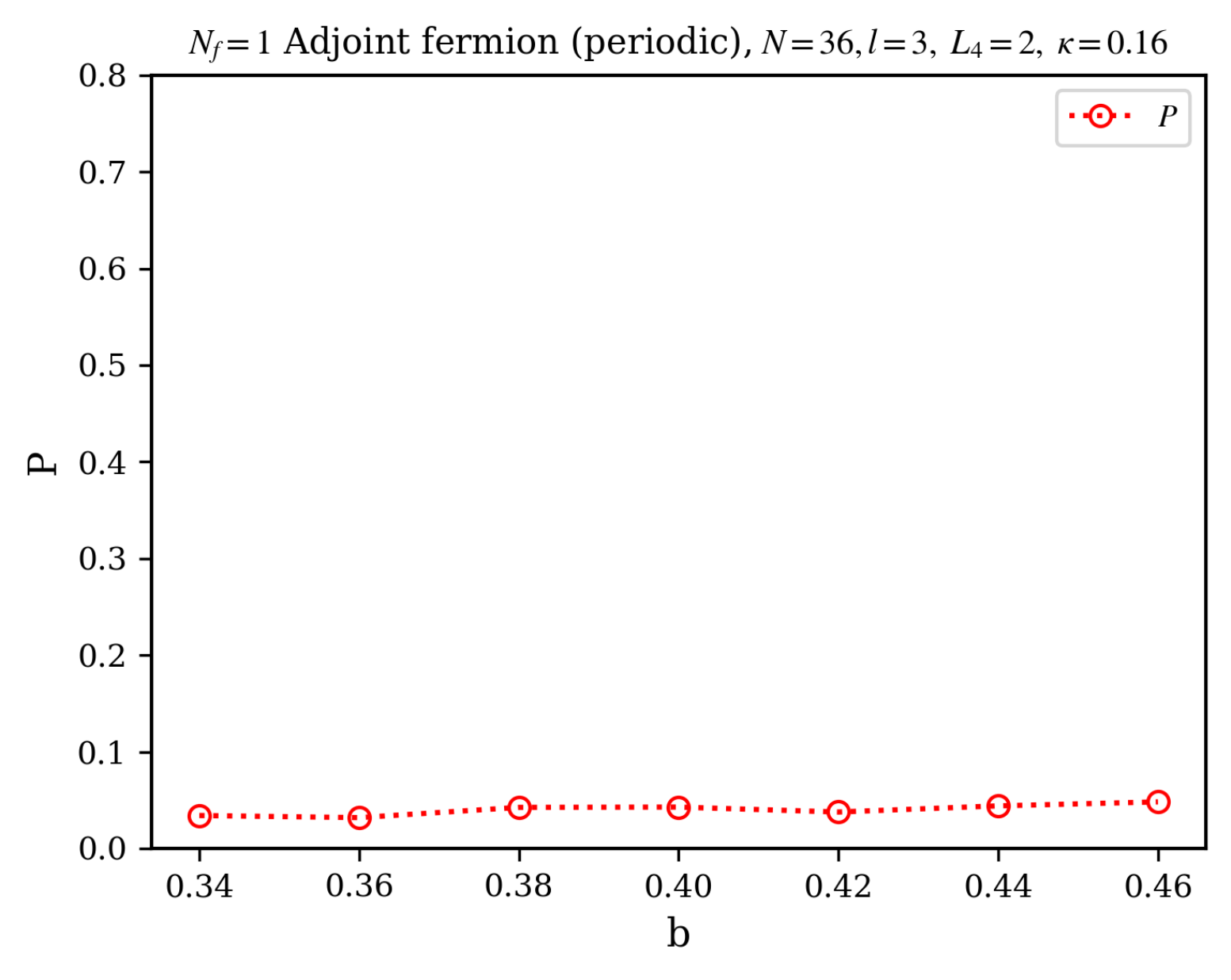}
\caption{
Deconfinement order parameter $P$ (red circles) in the partially reduced TEK model with {\bf the modified twist}, $N=36$, $N_f=1$ (periodic b.c.), $l=3$ and $L_4 = 2$ as a function of $b$. (Left) Heavy adjoint fermions $\kappa=0.03$: $P$ shows a sharp rise around $b=0.36$, indicating deconfining phase transition. (Right) Lighter adjoint fermions $\kappa=0.16$: $P$ remains near zero over the entire range, with no sign of a phase transition. Error bars are smaller than the symbol size.
}
\label{fig:MT}
\end{figure}

Compared to our previous results for $N_f = 2$ using the same parameter set \cite{Hamada:2025whg}, the deconfinement transition and the behavior of $P$ in the heavy mass region for $N_f = 1$ differ significantly:
For $N_f = 2$, the Polyakov loop fluctuates above the critical $b$ for $\kappa = 0.03$, suggesting the partial breaking of the ${\mathbb Z}_N$ center symmetry.
On the other hand, this is not what we observe for $N_{f}=1$: There is no fluctuation of the Polyakov loop above the critical $b$, indicating that there is no partial breaking, but just the full breaking of ${\mathbb Z}_N$.

A plausible interpretation is that this qualitative difference is attributed to the strength of the center-stabilizing effect induced by the adjoint fermions: Since adjoint fermions generate an effective potential that favors the $\mathbb{Z}_N$-symmetric vacuum, increasing the number of flavors $N_f$ enhances this stabilization. Therefore, it is highly plausible that for $N_f=2$, the stronger restoring force prevents the complete breakdown of the center symmetry even above the critical $b$, resulting in a partial breaking and the observed fluctuations of the Polyakov loop. In contrast, for $N_f=1$, the stabilization effect is weaker, leading to a direct transition to the fully broken phase without any intermediate partially-broken states.


\subsection{Schematic phase diagram for $N_{f} = 1$ QCD(adj.)}

In the partially reduced setup, we can summarize the phase structure in terms of confinement vs. deconfinement: For antiperiodic (thermal) boundary condition for a single adjoint quark, there is a line of deconfinement transitions in the $(\kappa,b)$ plane.  For a single periodic fermion, our data show no such transition in the light-mass region while we have it in the heavy-mass region. Since $N_f=1$ adjoint QCD is not expected to lie near the conformal window, our result indicates that the center-symmetric confined phase extends continuously from heavy to lighter masses within the range we studied.

\begin{figure}[t]
\centering
\includegraphics[width=10cm]{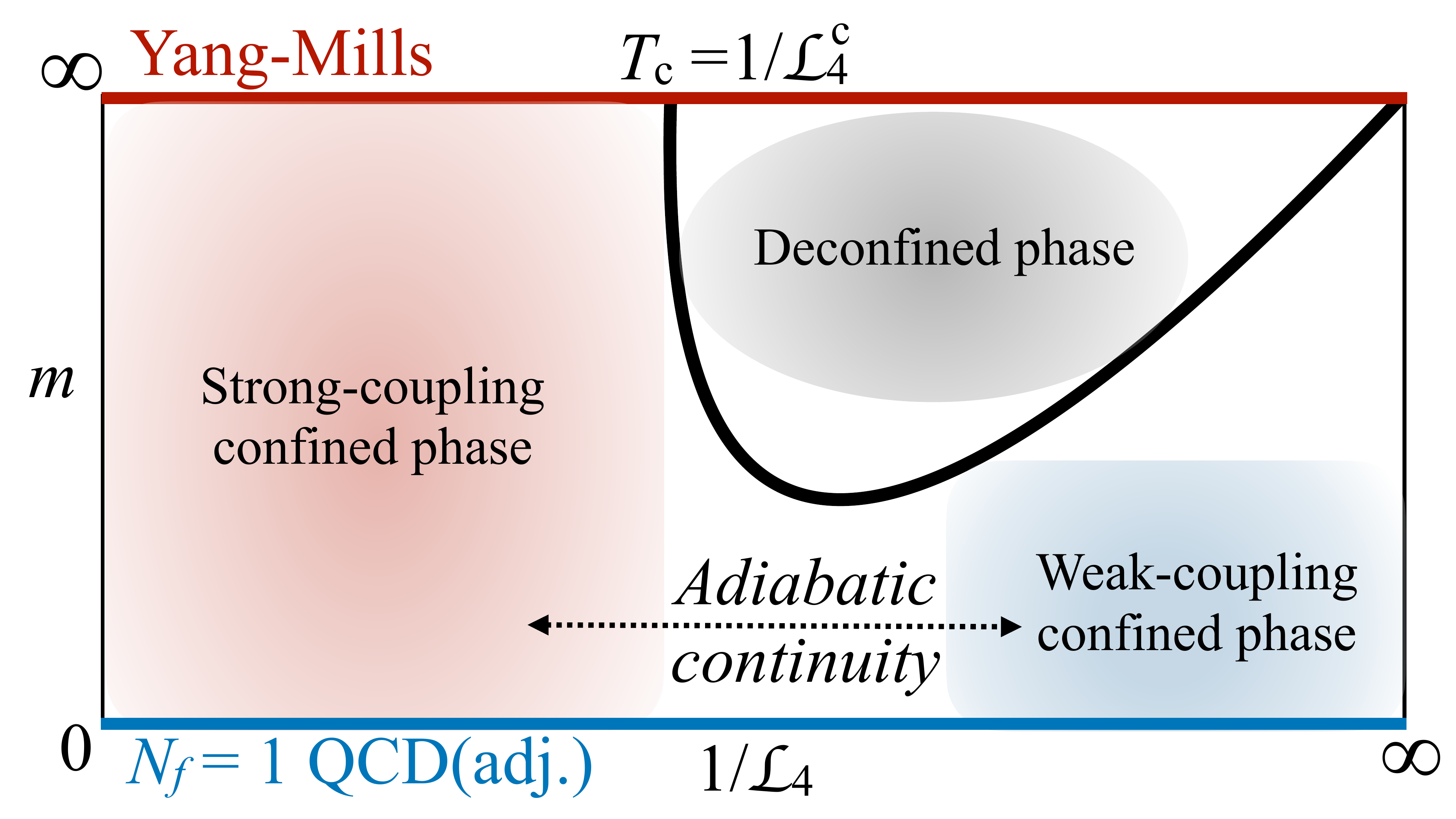}
\caption{Schematic illustration of confining adiabatic continuity scenario for $SU(N)$ gauge theory with $N_f=1$ periodic adjoint fermion on ${\mathbb R}^3 \times S^1 $ with the circumference ${\mathcal L}_{4}$. }
\label{fig:AC}
\end{figure}

Figure~\ref{fig:AC} schematically illustrates the phase structure for continuum $SU(N)$ gauge theory with $N_f=1$ periodic Dirac adjoint fermion on ${\mathbb R}^3 \times S^{1}$ with the circumference ${\mathcal L}_{4}$:
In pure Yang-Mills theory corresponding to the $m\to \infty$ limit, one encounters a critical point $T_c = 1 / {\mathcal L}_{4}^{c}$ as $1/{\mathcal L}_{4}$ increases, where the Polyakov loop in the $S^1$ direction gets non-zero. 
For light mass (small $m$), the system remains confined for all $1/{\mathcal L}_{4}$, where no phase transition separates large ${\mathcal L}_{4}$ confined phase (strong-coupling confined phase) from small ${\mathcal L}_{4}$ confined phase (weak-coupling confined phase). We note the latter confined phase is expected semi-classically to be described by the bion-confinement mechanism \cite{Unsal:2007vu, Unsal:2007jx,Kovtun:2007py, Shifman:2008ja, Misumi:2014raa, Poppitz:2021cxe, Hayashi:2024psa}. 
The numerical results shown in Figs.~\ref{fig:AC1} and \ref{fig:MT} are consistent with this schematic scenario of phase structure. 
In a more detailed study, one could attempt to determine the precise $\kappa_c$ $(m_c)$ at which adjoint quarks become light enough to affect the long-distance physics; we leave this for future work.


\section{'t Hooft anomaly and consistency of adiabatic continuity}
\label{sec:anomaly}

In this section, we discuss the relation between the adiabatic-continuity scenario and the mixed 't Hooft anomaly of one-flavor adjoint QCD. The purpose here is not to prove the absence of a phase transition from symmetry alone, but to show that the continuity scenario observed in Sec.~\ref{sec:AC} is fully consistent with the anomaly constraints, once the following assumptions are satisfied: (i) the partially reduced model employs the modified twist in Eq.~(\ref{eq:mtwist}), (ii) full volume independence holds, namely all Wilson lines $W(n_1,n_2,n_3)$ in Eq.~(\ref{eq:wn123}) vanish in the large-$N$ limit, and (iii) the theory is considered at the massless point $\kappa=\kappa_c$, where the discrete chiral symmetry of the continuum theory is restored.

For one Dirac fermion in the adjoint representation, the massless continuum theory has the one-form center symmetry $\mathbb{Z}_N^{[1]}$ and the discrete chiral symmetry $(\mathbb{Z}_{4N})_\chi$ (for example, see \cite{Misumi:2014raa}). The continuous axial $U(1)_A$ is broken by the ABJ anomaly, while the discrete subgroup $(\mathbb{Z}_{4N})_\chi$ remains exact. In the presence of the background two-form gauge field $B$ for $\mathbb{Z}_N^{[1]}$, the partition function transforms under the generator of $(\mathbb{Z}_{4N})_\chi$ as
\begin{equation}
Z[B]\;\longrightarrow\; \exp\!\left( -\frac{iN}{4\pi}\int_{M_4} B\wedge B \right) Z[B] \,,
\label{eq:disc-anomaly}
\end{equation}
where $M_4$ is a closed spin manifold.
This is the mixed center-chiral anomaly \cite{Gaiotto:2017yup, Shimizu:2017asf, Yonekura:2019vyz}. It is the same anomaly phase as in the $\mathcal{N}=1$ SYM case, although the discrete chiral group is enlarged from $\mathbb{Z}_{2N}$ to $\mathbb{Z}_{4N}$ for one Dirac adjoint fermion.

Upon compactification to the partially reduced setup, the parent four-dimensional one-form center symmetry $\mathbb{Z}_N^{[1]}$ gives rise to several lower-dimensional descendants. In the present $1^3\times L_4$ model, this includes the $(\mathbb{Z}_N)^3$ center symmetries acting on the three reduced directions and the $\mathbb{Z}_N$ center symmetry associated with the non-reduced compact $S^1$ direction. These descendants play different roles. The former control volume independence and specify the nontrivial flux sector selected by the twist as discussed in Sec.~\ref{sec:FT}, while the latter is the symmetry probed by the Polyakov loop in Eq.~(\ref{eq:acP}) and is directly relevant to the adiabatic continuity as investigated in Sec.~\ref{sec:AC}. Thus, while the mixed anomaly in Eq.~(\ref{eq:disc-anomaly}) is understood as a statement about the parent four-dimensional $\mathbb{Z}_N^{[1]}$ symmetry, it gives nontrivial implications for its descendants realized in different ways in the partially reduced model.

The relevance of this anomaly to the partially reduced model becomes transparent once the modified twist is interpreted as selecting a fixed nontrivial 't Hooft-flux sector in the three reduced directions. Under the full volume independence, the $1^3\times L_4$ model should be viewed as a reduced large-$N$ realization of adjoint QCD on an emergent $T^3\times S^1$ with  a fixed center background. 
Here the Hilbert-space discussion should be understood as a canonical reformulation of the same four-dimensional anomaly, obtained by quantizing the parent theory on spatial $T^3$ with an
additional time direction $\mathbb{R}$ \cite{Cox:2021vsa, Tanizaki:2022ngt}. By contrast, in the partially reduced model the compact direction of size $L_4$ is interpreted as a spatial circle, as appropriate for periodic adjoint fermions.

Denoting the 't Hooft flux by integers $n_{ij}$ $(i,j=1,2,3)$, one may introduce the magnetic-flux vector
\begin{equation}
m_k=\frac{1}{2}\,\epsilon_{kij}n_{ij}\, ,
\label{eq:magnetic-vector}
\end{equation}
which characterizes the twisted sector. In a canonical quantization on $T^3$, the generators $\hat T_j$ of the center symmetry and the generator $\hat X$ of the discrete chiral symmetry obey the centrally extended algebra \cite{Cox:2021vsa}
\begin{equation}
\hat T_j \hat X = e^{-\,2\pi i m_j/N}\, \hat X \hat T_j \, .
\label{eq:central-extension}
\end{equation}
Equivalently, if $|E,e_j\rangle$ denotes an energy eigenstate with electric-flux quantum number $e_j$, one finds
\begin{equation}
\hat X |E,e_j\rangle = |E,e_j-m_j\rangle \, .
\label{eq:shift-electric-flux}
\end{equation}
Thus the discrete chiral transformation cyclically permutes the electric-flux sectors. When at least one component of $\vec m$ is nonzero modulo $N$ and is coprime to $N$, repeated action of $\hat X$ generates $N$ distinct states. This implies an $N$-fold degeneracy and the spontaneous breaking pattern
\begin{equation}
(\mathbb{Z}_{4N})_\chi \;\longrightarrow\; \mathbb{Z}_4 \, ,
\label{eq:breaking-pattern}
\end{equation}
which is the natural anomaly-matching pattern for one-flavor adjoint QCD in a confining phase.

This observation has an important consequence for the small-circle (large-$b$) regime of the partially reduced model on $1^3\times L_4$. If the modified twist preserves full volume independence, then the reduced model should inherit the same anomaly constraints as the underlying four-dimensional theory in the corresponding nontrivial flux sector. Therefore, the small-circle regime cannot be a trivially gapped phase with both the center symmetry associated with the compact $S^1$ direction and the discrete chiral symmetry unbroken. In other words, the anomaly excludes a completely trivial center-symmetric and chirally symmetric vacuum.

At the same time, the anomaly constrains how the infrared theory must realize the symmetry. Hence the adiabatic-continuity scenario is perfectly compatible with the anomaly: the large-circle (small-$b$) strong-coupling confined phase and the small-circle (large-$b$) weak-coupling confined phase may belong to the same anomaly-matching class, 
so long as the center symmetry associated with the compact $S^1$ remains unbroken and the discrete chiral symmetry is realized nontrivially, for example through the breaking pattern in Eq.~(\ref{eq:breaking-pattern}). In this sense, the anomaly provides a nontrivial consistency check on the adiabatic-continuity scenario.

Our numerical results in Sec.~\ref{sec:AC} are consistent with this picture. In particular, for the modified twist we found that the volume-independence order parameters are consistent with zero, while the Polyakov loop $P$ in Eq.~(\ref{eq:acP}) also remains small in the light-fermion regime. The anomaly argument given above does not by itself prove $P=0$ or the absence of a phase transition. However, it shows that the adiabatically connected confined phase observed in the partially reduced model is not in conflict with the fundamental anomaly structure of the underlying four-dimensional one-flavor adjoint gauge theory.


\section{Summary and Discussion}
\label{sec:SD}

We have investigated, within a partially reduced twisted Eguchi-Kawai framework, whether the confined phase of one-flavor adjoint QCD on $\mathbb{R}^3 \times S^1$ can persist from large to small circle size under periodic compactification. To do so, we measured both the Polyakov loop around $S^1$ and the order parameters for volume independence in the reduced directions. Among the two twists we examined, the modified twist provides the cleaner setting: in the explored region, its volume-independence order parameters remain consistent with zero, whereas the symmetric twist shows signs of residual $({\mathbb Z}_N)^3$ symmetry breaking at $N=36$.

Within this setup, our numerical data in the light-fermion regime show no sign of a deconfinement transition for periodic boundary conditions as the circle is reduced, while antiperiodic boundary conditions lead to the expected thermal deconfinement transition. Taken together, these results provide numerical evidence, in the explored parameter region, for an adiabatic-continuity scenario in which the large-circle and small-circle confined phases are smoothly connected. Our anomaly discussion further shows that such a scenario is not in conflict with the symmetry constraints of the underlying four-dimensional theory.

In summary, the partially reduced adjoint TEK model with a modified twist and one-flavor light Dirac adjoint fermion with periodic boundary conditions yields results consistent with both volume independence and confinement throughout the parameter range studied. This supports, though does not by itself establish in a mathematically complete sense, the adiabatic continuity conjecture for large-$N$ gauge theories with adjoint matter.

For future work, we suggest the following directions:
\begin{itemize}

\item Extend the parameter scan to larger $b$ and $N$: Increase $b$ further to approach the continuum limit and check if the $(\mathbb{Z}_N)^3$-symmetric phase persists to $b\to\infty$.  Also, simulate larger $N$ to reduce $1/N$ corrections and solidify the volume equivalence.

\item Explore light-mass dynamics: Investigate the phase diagram at larger $\kappa$ to locate any onset of new behavior as the adjoint fermion becomes lighter.  

\item Consider $N_f=1/2$ (SYM): Another extension is to study the twisted reduction with a single Weyl (or Majorana) adjoint fermion, corresponding to $\mathcal{N}=1$ SYM. 

\item $SU(N)$ gauge theory with heavy adjoint and $N$-flavor light fundamental fermions on ${\mathbb R}^3 \times S^1$: Such a model has been investigated \cite{Cherman:2017tey}, where ${\mathbb Z}_N$-twisted boundary condition is imposed on the fundamental fermions in $S^1$ \cite{Kouno:2012zz, Sakai:2012ika, Kouno:2013zr, Kouno:2013mma, Kouno:2015sja, Iritani:2015ara, Misumi:2015hfa, Tanizaki:2017qhf, Tanizaki:2017mtm, Dunne:2018hog}. It is interesting to construct the TEK model and the partially reduced model corresponding to this setup.

\item Resurgent structure: If the adiabatic continuity holds, the resurgent structure, where the Borel-resummed perturbative series is non-trivially related to the non-perturbative contribution, is also expected to continue from weak-coupling to strong-coupling regimes \cite{Argyres:2012ka,Argyres:2012vv,Dunne:2012ae,Dunne:2012zk,Misumi:2014jua,Misumi:2014bsa,Misumi:2014rsa,Misumi:2015dua,Misumi:2016fno,Fujimori:2016ljw,Fujimori:2017oab,Fujimori:2017osz,Fujimori:2018kqp,Nishimura:2021lno,Sueishi:2020rug,Sueishi:2021xti,Kamata:2021jrs,Misumi:2024gtf,Misumi:2025ijd}. One may be able to investigate this issue by use of the lattice stochastic perturbation in the partially reduced model.
\end{itemize}

\begin{acknowledgments} 
The authors thank the YITP workshop ``Progress and Future of non-perturbative quantum field theory" (YITP-W-25-25) and the YITP long-term workshop ``Hadrons and Hadron Interactions in QCD 2024'' (YITP-T-24-02) for providing opportunities of useful discussions.  This work was partially supported by Japan Society for the Promotion of Science (JSPS) KAKENHI Grant No.~23K03425 (T.M.) and 22H05118 (T.M.).
The numerical calculations were partly carried out on Yukawa-21 at YITP in Kyoto University.
\end{acknowledgments}

\appendix

\section{Why is $W_{123}$ special?}
\label{sec:sol}

In this appendix, we clarify why $W_{123}$ in Eq.~(\ref{eq:w123}) plays a critical role in the investigation of volume independence. By examining the classical solutions, we demonstrate how volume independence can be broken in the partially reduced model with the symmetric twist defined in Eq.~(\ref{eq:twist}).

In the weak-coupling limit $b \rightarrow \infty$, the link variables $U_{\mu}(\tau)$ that minimize the action of the $1^3 \times L_4$ partially reduced TEK model satisfy the following classical conditions:
\begin{align}
    U_i(\tau) U_j(\tau) &= \exp \left(\frac{2 \pi i k}{L} \right) U_j(\tau) U_i(\tau), \quad (i < j) \label{con1} \\
    U_i(\tau + 1) &= U_4^{\dagger}(\tau) U_i(\tau) U_4(\tau)\,,
     \label{con2}
\end{align}
with $L^2 = N$ and $i,j=1,2,3$. 
The link variables satisfying the condition in Eq.~(\ref{con1}) in the reduced directions can be parameterized as
\begin{align}
    & U_1(\tau) = P_L \otimes D_1(\tau)\,,  \\
    & U_2(\tau) = Q_L \otimes D_2(\tau)\,,  \\
    & U_3(\tau) = Q_L P_L^{\dagger} \otimes D_3(\tau)\,,
\end{align}
where $P_L$ and $Q_L$ are the $L \times L$ shift and clock matrices satisfying the algebra:
\begin{equation}
    P_L Q_L =  \exp \left(\frac{2 \pi i k}{L} \right) Q_L P_L\,.
\end{equation}
Here, $D_i(\tau)$ are mutually commuting $L \times L$ matrices. 

From the condition in Eq.~(\ref{con2}), $U_4(\tau)$ can be expressed as
\begin{equation}
    U_4(\tau) = {\bm 1}_L \otimes V(\tau)\,,
\end{equation}
where $V(\tau)$ is an arbitrary $L \times L$ matrix. 
We now define a gauge transformation $\Omega(\tau) ={\bm 1}_{L}\otimes  V(1) V(2) \cdots V(\tau-1)$ with $\Omega(1) = \Omega(L_4 + 1) = {\bm 1}_N$. Under this transformation, $U_4(\tau)$ transforms as
\begin{equation}
\begin{split}
    U_4(\tau) \rightarrow \Omega(\tau) U_4(\tau) \Omega^{\dagger}(\tau + 1) &= \left({\bm 1}_L \otimes \prod_{s=1}^{\tau-1} V(s) \right) \left( {\bm 1}_L \otimes V(\tau) \right) \left({\bm 1}_L \otimes \prod_{s=1}^{\tau} V(s) \right)^{\dagger} \\ &= {\bm 1}_L \otimes {\bm 1}_L\,, 
    \label{U0_t}
\end{split}
\end{equation}
indicating that $U_{4}(\tau)$ can be gauged into the identity matrix ${\bm 1}_L \otimes {\bm 1}_L$ for $\tau=1,2,\dots,L_4 -1$. Regarding the final link variable $U_4(L_4)$, the periodic boundary condition yields
\begin{equation}
\begin{split}
    U_4(L_4)  & \rightarrow \Omega(L_4) U_4(L_4) \Omega^{\dagger}(L_4 + 1) \\ &= \left( {\bm 1}_L \otimes \prod_{s=1}^{L_4 - 1} V(s) \right) ( {\bm 1}_L \otimes V(L_4)) \\
    & = {\bm 1}_L \otimes \prod_{s=1}^{L_4} V(s)  \equiv P_4 \,,
    \label{U0_L4}
\end{split}
\end{equation}
where $P_4$ is identified with the Polyakov loop operator, which can be diagonalized. 

Consequently, we find that even if the elementary link variables are traceless ($\mathrm{Tr}\,U_{\mu} = 0$), the trace of the specific Wilson line ${\cal W}_{123}$, defined as
\begin{equation}
    {\cal W}_{123} \equiv U_1(\tau) U_2^{\dagger}(\tau) U_3(\tau) = {\bm 1}_L \otimes (D_1(\tau) D_2^{\dagger}(\tau) D_3(\tau))\,,
\end{equation}
is not guaranteed to vanish. This non-vanishing behavior occurs because the twist-induced matrices $P_L$ and $Q_L$ completely cancel out in this specific product, leaving the identity matrix in the first subspace. Furthermore, as shown in Eqs.~(\ref{U0_t}) and (\ref{U0_L4}), $U_4(\tau)$ commutes with ${\cal W}_{123}$, meaning there is no kinematic restriction that forces $\mathrm{Tr}\,{\cal W}_{123} = 0$. 

In contrast, for the model with the modified twist in Eq.~(\ref{eq:mtwist}), $U_1(\tau)$ and $U_3(\tau)$ do not commute with ${\cal W}_{123}$. This non-commutativity strictly enforces $\mathrm{Tr}\,{\cal W}_{123} = 0$ at the classical level. Therefore, $W_{123}$ serves as a sharp order parameter to detect the potential breaking of volume independence in the symmetric twist case, and to confirm its restoration in the modified twist case.

\section{Simulation setups}
\label{sec:SS}
In this paper, we use the standard Rational Hybrid Monte Carlo (RHMC) algorithm. The Hermitian Dirac operator is defined as $Q = D_w \gamma_5$, and the required fractional power of $Q^2$ is approximated by a rational function with $N_{shifts} = 20$. The resulting shifted linear systems are solved using the multi-shift conjugate gradient (CG) algorithm. 

Our stopping condition is as follows.
Let $r_i = b - (Q^2 + \beta_i) x_i$ be the residual for the $i\text{-th}$ shifted system, where $b$ is the source vector, $\beta_i$ is the shift parameter, and $x_i$ is the corresponding solution vector.
During the molecular dynamics evolution, we impose the stopping condition $|r_{ref}|^2/|b|^2 < 10^{-10}$, where $r_{ref}$ denotes the residue of the reference shifted system in the multi-shift CG solver. 
At the global Metropolis accept-reject step, we use the tighter condition $|r_{ref}|^2/|b|^2 < 10^{-15}$.
The step size in the molecular dynamics evolution and the number of time steps in one trajectory were chosen such that the acceptance rate exceeded 50\%. 
For all simulations, we generated 150 trajectories for each value of $\kappa$ at fixed b, and computed the expectation values of the Polyakov loop and Wilson line using the last 50 trajectories. 
Most simulations were performed using a desktop PC equipped with a 12th Gen Intel(R) Core(TM) i9-12900K 3.20GHz.

\bibliographystyle{utphys}
\bibliography{./QFT,./refs,./references}

\end{document}